\documentclass[a4paper,12pt, epsfig]{article}
\usepackage{epsfig}
\usepackage{amssymb}
\usepackage{amsfonts}
\usepackage{amsmath}

\newskip\humongous \humongous=0pt plus 1000pt minus 1000pt

\newif\ifdtup

\catcode`\@=11

\@addtoreset{equation}{section}
\def\theequation{\thesection.\arabic{equation}}

\def\@normalsize{\@setsize\normalsize{15pt}\xiipt\@xiipt
\abovedisplayskip 14pt plus3pt minus3pt%
\belowdisplayskip \abovedisplayskip
\abovedisplayshortskip \z@ plus3pt%
\belowdisplayshortskip 7pt plus3.5pt minus0pt}

\def\small{\@setsize\small{13.6pt}\xipt\@xipt
\abovedisplayskip 13pt plus3pt minus3pt%
\belowdisplayskip \abovedisplayskip
\abovedisplayshortskip \z@ plus3pt%
\belowdisplayshortskip 7pt plus3.5pt minus0pt
\def\@listi{\parsep 4.5pt plus 2pt minus 1pt
     \itemsep \parsep
     \topsep 9pt plus 3pt minus 3pt}}

\relax

\catcode`@=12

\catcode`\@=11

\def\section{\@startsection{section}{1}{\z@}{3.5ex plus 1ex minus
   .2ex}{2.3ex plus .2ex}{\large\bf}}

\def\thesection{\arabic{section}}
\def\thesubsection{\arabic{section}.\arabic{subsection}}

\def\appendix{\setcounter{section}{0}
 \def\thesection{Appendix \Alph{section}}
 \def\thesubsection{\Alph{section}.\arabic{subsection}}
 \def\theequation{\Alph{section}.\arabic{equation}}}

\def\SymBoxes#1#2#3#4{\newdimen\un@t \un@t#3%
\raisebox{#1}{\rule{#2\un@t}{#4}\hskip-#2\un@t
\@tempdimb\un@t \advance\@tempdimb by-#4\@tempcntb#2\relax%
\@whilenum{\@tempcntb>0}\do{
\rule{#4}{\un@t}\hskip\@tempdimb \advance\@tempcntb by\m@ne}%
\hskip-#2\un@t \rule[\un@t]{#2\un@t}{#4}%
\rule[\un@t]{#4}{#4}\hskip-#4
\rule{#4}{\un@t}}\hskip-#4}                

\begin{document}

\newcommand{\beq}{\begin{equation}}
\newcommand{\eeq}{\end{equation}}
\newcommand{\bea}{\begin{eqnarray}}
\newcommand{\eea}{\end{eqnarray}}
\newcommand{\beas}{\begin{eqnarray*}}
\newcommand{\eeas}{\end{eqnarray*}}
\newcommand{\defi}{\stackrel{\rm def}{=}}
\newcommand{\non}{\nonumber}
\newcommand{\bquo}{\begin{quote}}
\newcommand{\enqu}{\end{quote}}
\renewcommand{\(}{\begin{equation}}
\renewcommand{\)}{\end{equation}}
\def\IZ{{\mathbb Z}}
\def\IR{{\mathbb R}}
\def\IC{{\mathbb C}}
\def\IQ{{\mathbb Q}}
\def\IP{{\mathbb P}}

\def \eqn#1#2{\begin{equation}#2\label{#1}\end{equation}}
\def\de{\partial}
\def\Tr{ \hbox{\rm Tr}}
\def\H{ \hbox{\rm H}}
\def\HE{ \hbox{$\rm H^{even}$}}
\def\HO{ \hbox{$\rm H^{odd}$}}
\def\K{ \hbox{\rm K}}
\def\Im{ \hbox{\rm Im}}
\def\Ker{ \hbox{\rm Ker}}
\def\const{\hbox {\rm const.}}
\def\o{\over}
\def\im{\hbox{\rm Im}}
\def\re{\hbox{\rm Re}}
\def\bra{\langle}\def\ket{\rangle}
\def\Arg{\hbox {\rm Arg}}
\def\Re{\hbox {\rm Re}}
\def\Im{\hbox {\rm Im}}
\def\exo{\hbox {\rm exp}}
\def\diag{\hbox{\rm diag}}
\def\longvert{{\rule[-2mm]{0.1mm}{7mm}}\,}
\def\a{\alpha}
\def\dag{{}^{\dagger}}
\def\tq{{\widetilde q}}
\def\p{{}^{\prime}}
\def\W{W}
\def\N{{\cal N}}
\def\hsp{,\hspace{.7cm}}
\newcommand{\C}{\ensuremath{\mathbb C}}
\newcommand{\Z}{\ensuremath{\mathbb Z}}
\newcommand{\R}{\ensuremath{\mathbb R}}
\newcommand{\rp}{\ensuremath{\mathbb {RP}}}
\newcommand{\cp}{\ensuremath{\mathbb {CP}}}
\newcommand{\vac}{\ensuremath{|0\rangle}}
\newcommand{\vact}{\ensuremath{|00\rangle}}
\newcommand{\oc}{\ensuremath{\overline{c}}}
\begin{titlepage}
\begin{flushright}
ULB-TH/07-35\\
\end{flushright}
\bigskip
\def\thefootnote{\fnsymbol{footnote}}

\begin{center}
{\Large {\bf Gauge Theory RG Flows \\}}
\vspace{0.02em}
{\Large {\bf from \\}}
\vspace{0.6em}
{\Large {\bf a Warped Resolved Orbifold}} 
\\ 
\end{center}

\bigskip
\begin{center}
{\large  Chethan  
KRISHNAN\footnote{\texttt{Chethan.Krishnan@ulb.ac.be}} and Stanislav 
KUPERSTEIN\footnote{\texttt{skuperst@ulb.ac.be}}}\\
\end{center}

\renewcommand{\thefootnote}{\arabic{footnote}}

\begin{center}
\vspace{1em}
{\em  { International Solvay Institutes,\\
Physique Th\'eorique et Math\'ematique,\\
ULB C.P. 231, Universit\'e Libre
de Bruxelles, \\ B-1050, Bruxelles, Belgium\\}}

\end{center}

\noindent
\begin{center} {\bf Abstract} \end{center}
We study gauge-string duality of 
D3-branes localized at a point on the resolution of the ALE 
orbifold $\IC^3/\IZ_3$. By 
explicitly 
solving for the warp factor, we demonstrate the holographic RG flow from
$AdS_5 \times S^5/\IZ_3$ (far away from the resolution) 
to the usual 
$AdS_5\times S^5$ (close to the stack). On the gauge theory side, 
this maps to the flow between the quiver gauge theory and ${\cal N}=4$ 
super Yang-Mills. 
We present two possible scenarios for this RG flow depending on the choice of the VEVs.
In particular, one of the scenarios proceeds by two steps involving both
Higgsing and confinement. 

\vspace{1.6 cm}

\vfill

\end{titlepage}
\bigskip

\hfill{}
\bigskip

\tableofcontents

\setcounter{footnote}{0}
\section{\bf Introduction} \label{intro}

\noindent 
One powerful way of constructing gauge theories with reduced 
supersymmetry 
in string theory is to consider stacks of D-branes at Calabi-Yau 
singularities \cite{Klebanov:1998hh}. The original proposal of the 
AdS/CFT correspondence \cite{Maldacena:1997re} compared string theory on 
the near-horizon region of a stack of D3-branes in flat space to the low 
energy ${\cal N}=4$ super Yang-Mills theory living on the branes. The 
zoom-in involved in the near-horizon limit washes out all the details of 
the original background, so changing the background geometry 
from flat to curved does not result in new gauge theories.  
One way to get past this problem and to construct new gauge theories,
is to look at 
D3-branes not at a smooth point, but instead at the tip of a Calabi-Yau 
cone. If the cone $Y_6$ is a 6-dimensional real cone over a compact 
base space $X_5$, then the near-horizon limit gives rise to string 
theory on\footnote{The
way in which this happens is easily seen from, e.g., the discussion
we give in section 3.1.} 
$AdS_5 \times X_5$ and the dual gauge theory is generically a quiver 
theory whose details depend on the nature of the singularity.

To construct phenomenologically interesting theories, we need less 
supersymmetry and we also need to break conformal invariance. 
One way to break some of the SUSY is to replace the $S^5$ in $AdS_5 \times 
S^5$ with an 
orbifold $S^5/\Gamma$, where $\Gamma$ is a discrete group. This amounts 
to considering $\IR^{3,1}\times \IR^6/\Gamma$ instead of 
$\IR^{9,1}=\IR^{3,1}\times \IR^6$ before the near-horizon limit. 
If $\Gamma \in SU(2)$, then the dual 
theory is ${\cal N}=2$, and if $\Gamma \notin SU(2)$ then we have 
an ${\cal N}=1$ gauge theory\footnote{We still need $\Gamma \in SU(3)$, 
to satisfy the Calabi-Yau condition.}. In this paper, we will consider the 
case $\Gamma=\IZ_3$ and the dual ${\cal N}=1$ superconformal quiver. 
Another class of examples with reduced supersymmetry where the 
explicit Ricci-flat metric on the cone is known, is when the base 
$X_5$ is topologically $S^2\times S^3$. The most extensively investigated  
gauge-theory/singular-geometry dual of this kind is where $Y_6$ is chosen 
to be the conifold \cite{Candelas:1989js}, so that the base space is a 
coset space called $T^{1,1}$. More general classes of Ricci-flat 
metrics on cones over $S^2 \times S^3$ are also known, leading to the 
so-called $Y^{p,q}$ and (even more general) 
$L^{a,b,c}$ spaces and their ${\cal N}=1$ gauge theory duals \cite{
Gauntlett:2004yd, Bertolini:2004xf, Benvenuti:2004dy, Cvetic:2005ft,
Benvenuti:2005cz, Benvenuti:2005ja, Franco:2005sm, Cvetic:2005vk,
Martelli:2005wy, Berenstein:2005xa, Franco:2005zu, Bertolini:2005di,
Evslin:2007au, Martelli:2004wu, Feng:2000mi, HEK, Burrington:2005zd, 
Gepner:2005zt, Argurio:2006ny, Argurio:2007qk}. 

In \cite{Klebanov:1998hh, Morrison:1998cs}, the dual 
${\cal N}=1$ superconformal gauge theory 
corresponding to a stack of $N$ D3-branes at the tip of the conifold was 
argued to be an $SU(N)\times SU(N)$ gauge theory with matter content 
given by two 
chiral superfields $A_i$ and $B_i$ (where $i=1,2$). 
The fields $A_i$ transform in $(N,\bar N)$ of  $SU(N) \times SU(N)$ 
while $B_i$ transform in $(\bar N, N)$. The gauge theory on the singular 
conifold corresponds to the case where none of the fields have a VEV, and 
the theory is conformal. 

To break conformal invariance, we need to look at a space that is not  
a cone, because the radial direction of the cone is what gets 
absorbed into the $AdS^5$ when we take the near-horizon limit\footnote{The 
global symmetry $SO(4,2)$ of $AdS^5$ is what gets translated to the 
conformal symmetry of the gauge theory.}. In the case 
of the conifold theory described above, there are two ways in which we can 
accomplish this. One is to {\it deform} it by considering a non-vanishing 
$S^3$ at the tip, and the other is to {\it resolve} it and keep the $S^2$. 
The former case was the subject of the celebrated paper of Klebanov and 
Strassler \cite{Klebanov:2000hb} where the deformed $S^3$ is supported 
by a non-vanishing $3$-form flux. The dual theory was found to be a 
cascading gauge theory (related, but different, from the gauge theory on 
the singular space). The resolved conifold has also been 
investigated, but in the dual gauge theory, it corresponds to giving 
a non-zero VEV for the operator 
\eqn{BC}{\mathcal{U}=\frac{1}{N} {\rm Tr} \left( \left\vert A_1^2 
\right\vert + \left\vert A_2^2 \right\vert - 
                               \left\vert B_1^2 \right\vert - \left\vert B_2^2 \right\vert\right) }
in the {\it same} $SU(N) \times SU(N)$ gauge theory encountered in the 
case of the singular conifold. The dual 
supergravity solution under the assumption that the branes are smeared on 
the $S^2$ was constructed in \cite{Pando Zayas:2000sq}. The solution they 
found was singular, as expected. Ideally though, one would like to 
see the RG flow created by the specific non-zero VEVs on the supergravity 
side as well, and see the flow from $AdS_5 \times T^{1,1}$ in the UV to 
$AdS^5 \times S^5$ near the stack, when the stack is localized on a specific 
point on the resolution. This was accomplished in \cite{Klebanov:2007us}, 
where they found the precise form of the warp factor.  

In this paper we consider instead the space $\IC^3/\IZ^3$, which after the 
near-horizon limit gives rise to $AdS_5 \times S^5/\IZ_3$. We look at the 
case where the space gets a resolution, and the orbifold singularity 
is replaced by a four-cycle instead of the two-cycle of the conifold. 
This four-cycle is the projective space $\IC\IP^2$.
We consider the case where the stack of D-branes is localized on a point 
on the resolution. Far away from the resolution, we see that the warp 
factor tends to the unresolved case while close to the stack we see the 
emergence of the $AdS$ throat because the stack is no longer at a 
singular point. 
Exactly like in the conifold case there is a single parameter in the resolved metric,
which controls the size of the ``blown-up" cycle. 
In the dual gauge theory, this 
results in a non-zero VEV for 
a six-dimensional, yet to be constructed \cite{Benvenuti:2005qb} operator. 
On the gauge theory side it corresponds to the fact that all baryonic symmetries 
are anomalous and no simple current of the form (\ref{BC}) is preserved.
It means, that unlike in the conifold case there is no straight-forward way to figure out
what fields acquire VEVs.
However, using the relation between the holomorphic coordinates on the orbifold and
the chiral superfields of the gauge theory we found that there are only two non-identical 
patterns for the RG flow triggered by different VEVs of the fields.
In both cases the VEVs breaks the 
conformal invariance and launches an RG flow
that yields the ${\cal N}=4$ SYM, but one of the two scenarios is more complicated then the other
and involves two 
steps involving Higgsing and confinement, the details of which are 
presented in Section \ref{five}.

The structure of this paper is as follows. In the next section we explain
the geometry of the orbifold and its resolution. In Section \ref{gastr}
we explain the supergravity solution corresponding to D3-branes on the 
unresolved space, and the corresponding dual quiver gauge theory. The next 
two sections are dedicated to the construction of the solution in the 
resolved space and a discussion of the RG flow on both the gravity and 
gauge theory sides of the duality. We end in the final section with some 
discussions and comments. Some of the technical details can be found in 
the appendices.

Various other aspects of the $\IC^3/\IZ_3$ orbifold have been studied in
\cite{Gukov:1998kn, Ganor:2002ae, Ezhuthachan:2006gu, Butti:2007jv}. Other 
recent papers which are of relevance to our work are \cite{Chen:2007em, 
Singh:2007vw, Cvetic:2007nv, Martelli:2007mk, Klebanov:2007cx}.

\section{The Toric Description of $\IC^3/\IZ_3$ and its Resolution}

The space $\IC^3/\IZ_3$ is defined as the three-dimensional complex space
$\IC^3$
under the identification
\eqn{orbifold}
{ \{ w_1, w_2, w_3 \} = \{ \eta w_1,\eta w_2, \eta w_3 \}, \ {\rm with} \
\eta^3=1.}
Clearly, the only orbifold point is the origin. The orbifold inherits the 
metric
\eqn{orb}{ds_6^2=dr^2+r^2d\Omega_{S^5/\IZ_3}^2}
from the flat metric on $\IC^3$. The subscript on the angular part is
supposed to indicate that the periodicities of the angles are of the
orbifold, not of flat space.

There is another, more algebraic, way in which we could define the above 
orbifold. This turns out to be very useful for comparisons with the 
gauge theory, so we explain it now. The idea is that the orbifold is fully 
defined by the 
invariant (under the orbifold action) monomials that one can construct 
from the coordinates $w_i$. There are ten such monomials as is easily 
checked, and the domain over which these collection of 
monomials is well-defined is precisely 
$\IC^3/\IZ_3$. So one could also define our orbifold by
\eqn{algebra}{\IC^3/\IZ_3=\IC[w_1^3, \, w_1^2 w_2, \, w_1^2 w_3, \, 
w_2^3, \, w_2^2w_1, 
w_2^2w_3, \, w_3^3, \, w_3^2w_1, \, w_3^2w_2, \, w_1w_2w_3].}
If we call these monomials $U_{ijk}$, with 
indices ${i,j,k}={1,2,3}$, then the algebraic 
relations between $U_{ijk}$ (e.g., $U_{112} U_{123}=U_{223} U_{111}$) 
captures the geometry of 
the orbifold. It should be understood that $U_{ijk}$ with different 
permutations of $i, j, k$ are identified. So we could also write,
\eqn{algebraGEN}{\IC^3/\IZ_3=\IC[U_{ijk}]/\{ {\rm algebraic \ relations \ 
between \ the }\ U_{ijk}{\rm's} 
\}. }

\begin{figure}
\begin{center}
\includegraphics[width=0.9\textwidth]{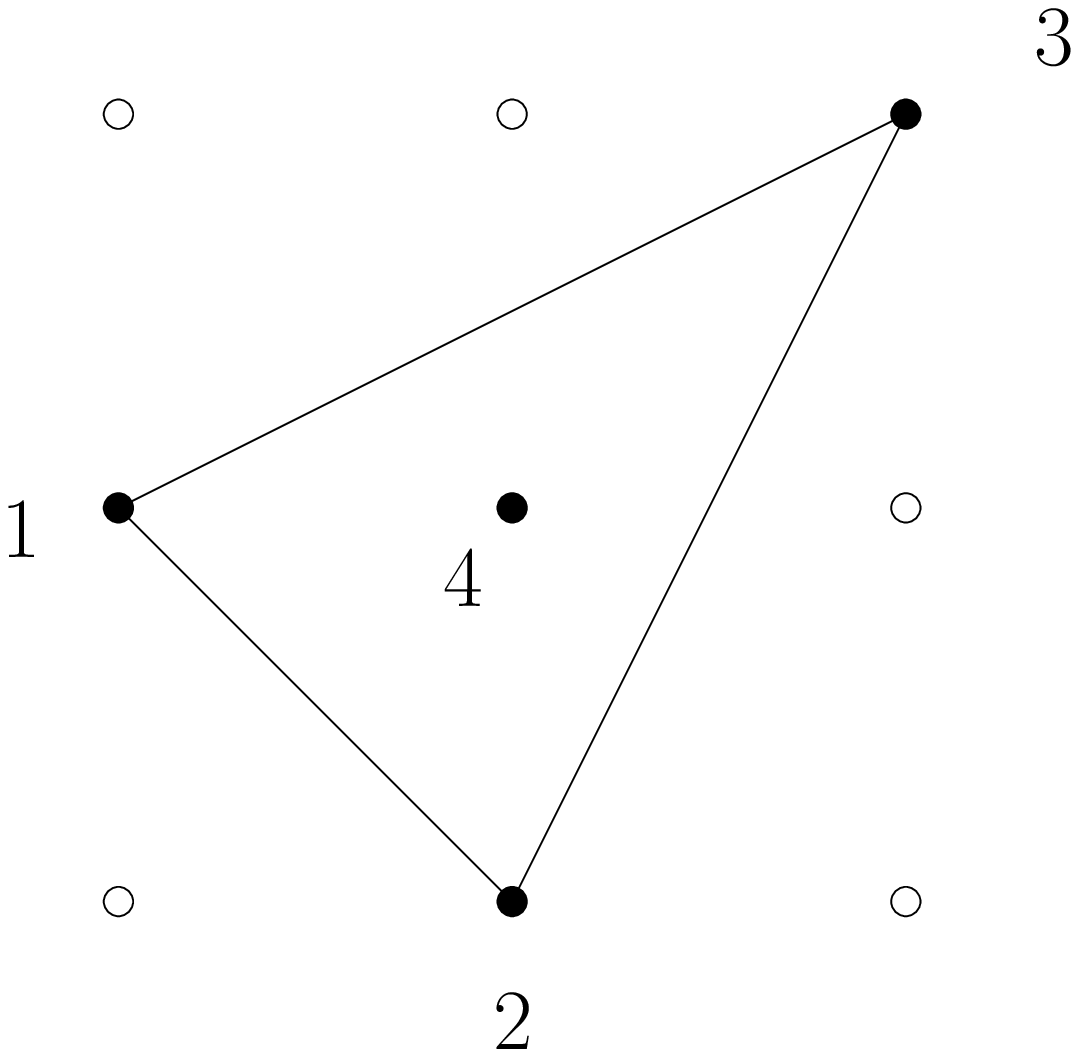}
\caption{Toric Diagram of $\IC^3/\IZ_3$.}
\label{toric}
\end{center}
\end{figure}

Another important useful fact about $\IC_3/\IZ_3$ is that it is  
toric. A quick and practical introduction to toric geometry can be 
found in appendix B of \cite{Argurio:2006ny} or section 4 of 
\cite{Bouchard:2007ik}.
Essentially, in the present context {\em toric} means that all the 
information about the space (as a complex variety) can be captured using 
a cone that has its apex at the origin of a 3-dimensional integer lattice. 
There are qualifications that need to be added to the cone to make this 
more precise, 
but instead of stating them, we will refer the exacting reader to
\cite{Bouchard:2007ik, Hori:2003ic, Fulton}. In any event, it turns 
out\footnote{See appendix 
B.1 of \cite{Ezhuthachan:2006gu} for a straightforward algorithm for 
reconstructing the space from its toric diagram.} 
that the cone that defines our orbifold can be specified by the 
(rational) vertices that span it: 
\eqn{vertex}{v_1=(-1,0,1), \ \ v_2=(0,-1,1), \ \ v_3=(1,1,1).} 
One feature of toric spaces that are in addition Calabi-Yau is that their 
vertices (other that the origin) all have to lie on the same plane. Thus, 
we can capture them on 
the toric diagram presented in Fig. \ref{toric}.

Toric diagrams are useful because many questions about the 
original space can be answered in terms of simple geometrical features of 
the diagram. The Calabi-Yau property being translated into 
co-planarity of vertices is a simple example. Another question that is 
readily answered is the question of singularities and their 
resolutions. In fact, the way in which the fact that 
$\IC^3/\IZ_3$ is singular is captured by our toric diagram is through the 
presence of a vertex, marked 4 in figure \ref{toric}, in its interior. 
In general, if the volume of the cone with origin as 
its apex is greater than one, then the 
space is singular. Notice that the volume can be greater than one 
only\footnote{We are talking about triangulated toric diagrams. The 
toric diagram for the conifold, for example, is a square with no interior 
points, but still the variety is singular. There the resolution 
corresponds to 
triangulating the square, which amounts to blowing up a 2-sphere, not 
a 4-cycle.} if 
there is an interior\footnote{By ``interior" here, we also mean 
points that are {\it on} a boundary. When the point is a genuine interior 
point like in our case, the resolution corresponds to the blowing-up of a 
four-cycle.} 
point (which in our case is the vertex marked 4). This motivates the 
following simple recipe for 
resolving our singular space: add more vertices in the interior in such a 
way that the new 
cones have no vertices in {\em their} interior, and then look at the 
resulting collection of cones (the ``fan") as the new space. In our case, 
since there is only one interior point, $\IC^3/\IZ_3$ has a unique 
(``crepant")
resolution, and the fan for the resolved space will involve the three new 
cones that have been created by the addition of vertex 4. 

We will be interested in the resolution of the orbifold in later sections, 
so we make some comments about that here. The vector for vertex 4 is easily 
seen to be
\eqn{v4}{v_4=(0,0,1).}
Clearly, the four vertices should satisfy 4 (= no. of 
vertices) - 3 (= dimensionality of the 
lattice) = 1 linear relation between 
them, which we write as 
\eqn{charge}{\sum_{i=1}^4 Q_i v_i =0, \ {\rm with} \ Q_i=(1,1,1,-3).}
The reason why these charges $Q_i$ are interesting is because 
the translation between the toric diagram and the actual space we are 
interested in, 
is effected through them. This is done by means of the ``quotient 
construction"
as follows. 
Pick 4 (= no. of vertices) complex coordinates $z_i$. We can define a 
$\IC^*$-action 
on these coordinates
\eqn{weighted}{(z_1, z_2, z_3, z_4) \rightarrow (\lambda z_1, \lambda z_2, 
\lambda z_3, \lambda^{-3} z_4), \ \lambda \in \IC^*,}
where the powers are the charges above. At this stage, we are only defining the action, we have not
yet modded out by anything. 
The space now is constructed by first imposing
\eqn{holo1}{|z_1|^2+|z_2|^2+|z_3|^2-3|z_4|^2=\mu,}
and then modding out by the phase part (the $U(1)$'s) of (\ref{weighted}). This should 
be compared to the two-step construction of $\IC\IP^1$ by 
first setting 
$|z_1|^2+|z_2|^2=\mu$ in $\IC^2$, and then modding out by phases, instead 
of modding out by $\IC^*$ in one go. Notice that the charges $Q$ affect 
both steps of the quotienting operation (\ref{holo1})-(\ref{weighted}).
This quotient construction of the geometry finds its gauge theory 
analogue in the description of the moduli space in terms of 
D-terms and is the basis for Witten's Gauged Linear Sigma Model (GLSM) 
construction \cite{Witten:1993yc} applied to ${\cal N}=1$ 
theories. 

Now, we can try to see how this toric construction of (the resolution of) 
$\IC^3/\IZ_3$ ties up with the earlier definitions. First we notice that 
instead of doing the $U(1)$ quotienting, we can parametrise the space
through the zero-charge monomials defined by
\eqn{homog}{z_1^3z_4, \, z_1^2 z_2z_4, \, z_1^2 z_3z_4, \, z_2^3z_4, \, z_2^2z_1z_4, 
z_2^2z_3z_4, \, z_3^3z_4, \, z_3^2z_1z_4, \, z_3^2z_2z_4, \, z_1z_2z_3z_4.}
Obviously, these monomials  satisfy the same algebra 
as the $U_{ijk}$'s in (\ref{algebra}). 

We claim that $\mu=0$ corresponds to  
the unresolved case, and when $\mu>0$ we get the resolved space where the 
origin is replaced by a $\IC\IP^2$. For this, first note that 
when $z_4=0$, all the monomials above are zero, which corresponds to the 
orbifold point in the language of the $U_{ijk}$'s. But from (\ref{holo1}) 
it is clear that when $\mu=0$, $z_4=0$ forces $z_{1,2,3}$ to be zero and 
therefore we end up with a single point, but when $\mu >0$, setting 
$z_4=0$ results in a $\IC\IP^2$. Thus, $\mu$ can be thought of as the size 
of the $\IC\IP^2$ on the resolved orbifold, and when it is zero we end up 
with $\IC^3/\IZ_3$. It is also possible to see the resolution using only 
the monomials
$U_{ijk}$ similar to the famous conifold construction of 
\cite{Candelas:1989js}. 
It follows from the algebra satisfied by $U_{ijk}$'s that \emph{all} $3$d 
complex vectors
of the form
\eqn{}{v_{jk}^T \equiv  \left( U_{1jk},U_{2jk},U_{3jk} \right)}
lie on the same $2$d plane,  which means that there exists a $3$d vector 
$\lambda$ we have
\eqn{vTlambda}{v_{jk}^T \cdot \lambda =0 \qquad  {\rm for  \,\,  any} 
\quad j,k .} This fact follows directly from relations of the form 
$U_{1jk} U_{2j^\prime k^\prime}=U_{2jk} U_{1j^\prime k^\prime}$.
The existence of $\lambda$ becomes even more evident if we notice that 
$v_{jk}^T=z_j z_k (z_1,z_2,z_3)$
for arbitrary $j$ and $k$. Now, the non-zero vector $\lambda$ is fixed up 
to an overall re-scaling, which means that it defines
a point on $\IC\IP^2$. Furthermore,  $\lambda$ is well-defined everywhere 
except the origin, where \emph{all}
the vectors $v_{jk}$ are identically zero. At the apex $\lambda$ is 
however un-restricted and we
resolve  the orbifold by replacing the singular point with the 
$\lambda$-parametrized $\IC\IP^2$. This is
exactly the same $\IC\IP^2$ we found setting $\mu > 0$.

Before we leave the geometry, we add some comments about how 
Higgsing in the gauge theory is described in the geometry. When $\mu > 0$ 
and $z_4=0$ the rest of the coordinates
in (\ref{holo1}) define $\IC\IP^2$. Let us assume that $\left \vert z_3 
\right \vert$ is of the same order
of magnitude as the parameter $\mu$. It means that we consider the 
$z_3\neq 0$ patch of the $\IC\IP^2$. On this patch we can introduce the 
coordinates $u_1\equiv z_1/z_3$ and $u_2\equiv z_2/z_3$. Together with 
the monomial $U_{333}=z_3^3 z_4$ these coordinates parametrise the entire 
patch $z_3 \neq 0$. Indeed, starting from the set $(u_1,u_2, U_{333})$ we 
can easily determine the rest of the $U_{ijk}$'s. For example, we have  
$U_{123}=u_1 u_2 U_{333}$. If we now send $\mu \to \infty$ still focusing 
on the $z_3 \neq 0$ patch we arrive at the regular $\IC^3$ parametrized by 
$u_1$, $u_2$ and $U_{333}$. On the toric diagram it looks like 
we ``chop" off the 3d node, while in the dual gauge theory it corresponds 
to the RG flow to the ${\cal N}=4$ SYM theory. 

\section{\bf Gauge-String Duality on the Unresolved Orbifold}
\label{gastr}

\subsection{Supergravity with Brane Sources}
\label{subra}

When we put a stack of D3-branes in some background, they act as sources 
for the type IIB supergravity equations of motion. There is a standard 
ans\"atz for solving these equations of motion (see e.g. 
\cite{Klebanov:1999tb}), which is given by the 
following prescriptions 
for the various fields (the metric, the dilaton and the five-form):
\begin{eqnarray}
ds^2=H^{-1/2}(y)\eta_{\mu\nu}dx^{\mu}dx^{\nu}+H^{1/2}(y)ds_6^2, 
\hspace{0.3in} \\
\Phi={\rm const}., \  \ F_5=(1+*)dH^{-1}\wedge dx^0 \wedge ... \wedge 
dx^3,
\end{eqnarray}
The $*$ stands for the Hodge dual operator. The $ds_6^2$ piece in the 
metric denotes the dimensions transverse to 
the D3-branes, and is given in our case by (\ref{orb}). The worldvolume 
(3+1)-metric of the D-branes is Minkowskian, and the entire solution is 
essentially given by one function, $H(y)$, where $y$ denotes the 
coordinates on the transverse space. This function is called the warp 
factor. With this ans\"atz, the supergravity EOMs (with source terms 
for the branes) reduce to a single 
equation, the Green's equation on the transverse space ($y_0$ denotes the 
location of the stack):
\eqn{gr}{\Box_y H(y,y_0)=-\frac{C}{\sqrt{g_6}}\delta^6(y-y_0).}
We denote the determinant of the 6-metric by $g_6$. The 
strength of the source is captured by $C=2 
\kappa_{10}^2T_3N=(2\pi)^4g_s 
N\alpha'^2$ where $T_3=1/(8\pi^3 g_s \alpha'^2)$ is the brane tension 
and $\kappa_{10}=8\pi^{7/2}g_s\alpha'^2$ is the 10-D Newton's constant. 

For the case of the unresolved $\IC^3/\IZ_3$, when we place the stack at 
the orbifold singularity, this equation can be immediately solved because 
the warp factor depends only on the radial coordinate. Green's 
equation takes the form 
\eqn{Gr1}{\frac{1}{r^5}\frac{\partial}{\partial 
r}\Big(r^5\frac{\partial}{\partial r} H\Big)= -\frac{3C}{\pi^3 
r^5}\delta(r).}
The normalization of the delta function accounts for the fact that we are 
ignoring the angular dependence, it comes from an integral over the 
angles\footnote{This is analogous to the fact that in 3-dimensions, if we 
are looking at sources at the origin ($r_0=0$), then the replacement 
$\frac{1}{r^2\sin 
\theta}\delta(r-r_0)\delta(\theta-\theta_0)\delta(\phi-\phi_0) 
\rightarrow \frac{1}{4\pi r^2} \delta(r)$
is acceptable for test-functions that are sufficiently well-behaved at 
the origin. The $4 \pi$ 
that arises is nothing but $\int_0^{\pi} \sin \theta 
d\theta\int_0^{2\pi} 
d\phi$.}. Away from 
the origin the equation is easily integrated, and integrating over the 
delta function fixes the constant of integration:
\eqn{warp1}{H(r)=\frac{L^4}{r^4} \ \ {\rm where} \  \ L^4=12 \pi g_s N 
\alpha'^2.}
With the usual substitution $z=L^2/r$, we end up with
\eqn{S5z3}{ds^2= 
\frac{L^2}{z^2}(dz^2+\eta_{\mu\nu}dx^{\mu}dx^{\nu})+L^2d\Omega_{S^5/\IZ_3}}
which is nothing but $AdS^5\times S^5/\IZ_3$ with equal radii ($=L$) for 
both the AdS and the $S^5/\IZ_3$. 

Another thing that we could do is to put the stack away from the tip, in 
which case we expect that far away, the solution should still look like 
the one we found above. But close to the stack, now we should see the 
emergence of the AdS throat because the stack is now at a smooth point. 
The full solution will have a singularity at $r=0$. We verify these 
expectations explicitly in an appendix. 

The general picture is that when the stack is at the singularity we have 
the $AdS^5\times S^5/\IZ_3$ background, and this corresponds to zero VEVs 
for the fields in the dual gauge theory. When we resolve the space, that 
is equivalent to turning on specific VEVs and one of the purposes of this 
paper is to investigate the RG flows triggered by these VEVs.  

\subsection{The Dual Quiver Theory}
\label{duq}

There are more or less standard recipes for constructing 
dual gauge theories corresponding to D-branes probing singularities. 
Building on the 
work of \cite{Douglas:1996sw, Douglas:1997de}, algorithms for constructing 
the dual gauge theory corresponding to generic toric singularities have 
been developed in a sequence of papers  starting with 
\cite{Feng:2000mi}. The state of the 
art can be found in the excellent review by Kennaway 
\cite{Kennaway:2007tq}.

\begin{figure}
\begin{center}
\includegraphics[width=0.7\textwidth]{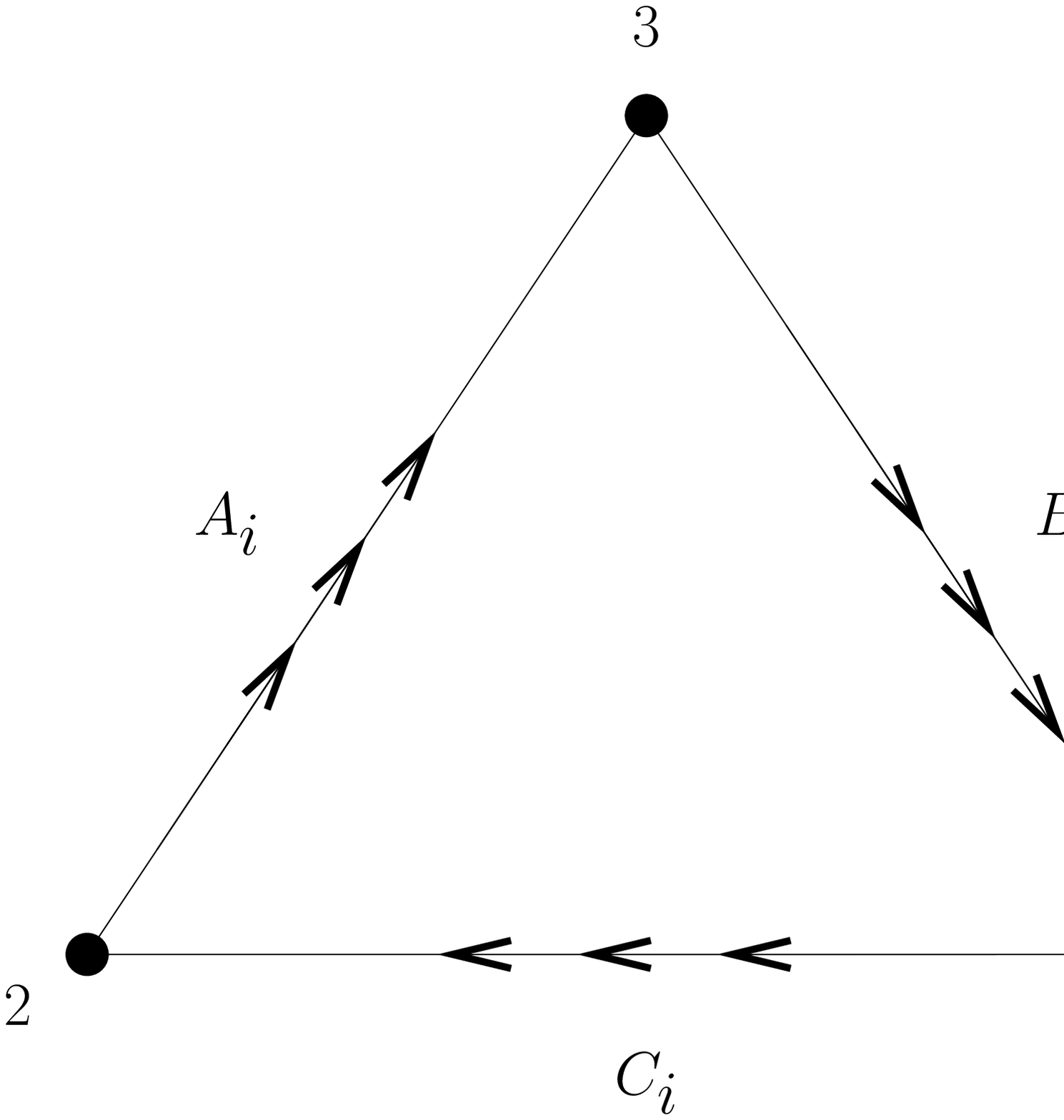}
\caption{The quiver for the gauge theory dual to $\IC^3/\IZ_3$.}
\label{Quiver1}
\end{center}
\end{figure}

But for Abelian orbifolds, like the one we are considering here, we 
can deduce the gauge theory from simple arguments starting with AdS/CFT 
duality for $AdS_5\times S^5$ \cite{Kachru:1998ys, Bigazzi:2003ui}.
It turns out that the gauge theory information can be captured using a 
simple quiver 
diagram (Fig. (\ref{Quiver1})). 
Quiver diagrams are merely a compact way of describing the field content 
in an ${\cal N}=1$ theory. In our case, there are three gauge groups 
and the fields transform as bi-fundamentals. For example,  $A_i 
\ (i=1,2,3)$,  transforms in the fundamental $N$ of $U(N)_3$ and 
and the anti-fundamental ${\bar N}$ of $U(N)_2$ in the figure. Notice that 
we have 
three sets of three fields each, one set on each edge. We also  assign 
charges for these fields under the $U(1)$ contained in the $U(N)$: fields 
in the fundamental of $U(N)$ are assigned a $+1$ under the associated 
$U(1)$ and those in the anti-fundamental are assigned a $-1$. For example, 
the $A_i$ have a charge of $-1$ under $U(1)_2$.

Apart from the $U(1)_R$ symmetry and the $SU(3)$ symmetry that acts on the $i$ indices,
there are no anomaly-free global continuous symmetries in the theory. In particular, there are two
candidates for the baryonic symmetry, but both prove to be anomalous as one can easily check.
As we have already mentioned in the Introduction this fact is evident from the geometry, since the
dual singular space has no two-cycle resolution, and so there is no parameter dual to the baryonic current. 
On the other hand, the theory enjoys some un-broken discrete symmetries. The precise form of these symmetries as well
their action on various wrapped $D$-branes on the geometry side is beautifully explained in 
\cite{Gukov:1998kn}.

For all of the nodes one has $N_{\rm f}= 3 N_{\rm c}$, 
where $N_{\rm f}$ denotes the number of the flavors and $N_{\rm c}=N$ corresponds to the colour.
Thus the $\beta$-functions vanish precisely
and the theory is conformal.

The quiver also determines the D-term equations that constrain the moduli 
space of vacua of the gauge theory. 
For the $U(1)$ on the $a$-th node with Fayet-Iliopoulos 
parameter $t_a$, it takes the form $\sum_m 
Q_{a,m}|X_m|^2=t_a$. The $Q$ here are the charges, the $X$ are the
fields, and the summation is over all fields. For our case then, 
\begin{eqnarray} \label{FI}
\sum_i (|A_i|^2-|B_i|^2)=t_3, \\
\sum_i (|B_i|^2-|C_i|^2)=t_1, \\
\sum_i (|C_i|^2-|A_i|^2)=t_2. \label{FI3}
\end{eqnarray}
Here we assume for simplicty that the branes are moving together, which 
means that the fields $A, B, C$ are assumed to be proportional to the 
identity. In the more general case, we will need to consider many copies 
of the space corresponding to independent motions of the branes. In 
any event, the fact that $t_1+t_2+t_3$ is identically zero, suggests that 
there are 
only two independent parameters instead of three: one corresponds to the 
resolution parameter $a$, and the other corresponds to the $B$ field that 
can be turned on in the background \cite{Benvenuti:2005qb}.

The D-terms do not fully fix the vacuum moduli space as can be clearly 
seen by counting the degrees of freedom (DOF). There are 3 $\times$ 3 
complex fields, and each D-term equation together with $U(1)$ condition 
constrains one DOF each. But 
we also have to make allowance for the fact that the D-terms for all the 
gauge groups add up to zero from the equations above, so they are not all 
independent. Thus the vacuum moduli space according to the D-terms alone 
is $3\times3-3+1=7$ complex dimensional, in conflict with our expectation 
that it should in fact be 3-dimensional. The resolution is of course that 
we haven't taken the F-terms into account yet, for which we need the 
superpotential.

The superpotential for the quiver is 
\eqn{sup}{W=\epsilon_{ijk} {\rm Tr}\left(A_iB_jC_k \right).}
The trace is over the gauge indices, and we will suppress writing it from 
now on. 
The F-term equations 
$\frac{\partial W}{\partial X}=0$ (where $X$ is $A_i, B_i \ {\rm or \ } C_i$)
therefore give rise to a bunch of algebraic relations of the form 
$B_2C_3= C_2B_3$ (this particular one arises for the choice $X=A_1$). 
The vacuum moduli space should be described by the
independent gauge-invariant chiral operators, modulo the F-term constraints.
The gauge-invariant chiral mesonic operators are of form $A_i B_j C_k$, 
and after imposing the F-term conditions the ones that are left are ten 
in number:
\begin{equation}  \label{MO}
\begin{array}{c}
A_1B_1C_1, A_1B_1C_2, A_1B_1C_3,  A_2B_2C_2, A_1B_2C_2, A_2B_2C_3,  \\
A_3B_3C_3,  A_1B_3C_3,  A_2B_2C_3, A_1B_2C_3.
\end{array}
\end{equation}
But now, it can immediately be checked that these have a one-to-one 
correspondence with
the ten monomials $U_{ijk}$ that we introduced in defining the orbifold 
in section 2, and
that the algebraic relations that they satisfy are precisely the 
same\footnote{Roughly speaking we find that the F-terms identify $z_i$ with 
the fields $(A_i,B_i,C_i)$.}. 
Therefore, the moduli
space is precisely the orbifold $\IC^3/\IZ_3$.

\section{\bf Branes on the Resolved Orbifold}
\label{resolvedbrane}

We saw in Section \ref{gastr} that the orbifold singularity of our 
orbifold
can be ``blown-up" by replacing it with the compact projective space 
$\IC \IP^2$ so that one ends up with a non-compact Calabi-Yau manifold 
that looks asymptotically like $\IC^3/\IZ_3$. We will be interested in 
doing supergravity with D3-brane sources localized on this resolved space. 
In particular, we want the explicit form of the metric on it  so that we 
can compute the Laplacian and solve for the 
warp factor in the full supergravity solution. 

A strategy for the derivation of this metric (which generalizes readily to 
other $\IC^n/\IZ_n$ spaces \cite{Nibbelink:2007rd, Ganor:2002ae}) is given 
in an 
Appendix. In terms of the angular variables defined by
\begin{eqnarray}
\label{eq:w1w2w3}
w_1&=& r \sin \sigma \sin \frac{\theta}{2}\ e^{ i 
\big(\frac{\psi}{3}+\frac{\phi}{2}-\frac{\beta}{2}\big)}, \\
w_2&=& r \sin \sigma \cos \frac{\theta}{2}\ e^{ i
\big(\frac{\psi}{3}-\frac{\phi}{2}-\frac{\beta}{2}\big)}, \\
w_3&=& r \cos \sigma \ e^{i\frac{\psi}{3}},
\end{eqnarray}
the metric takes the form (with $\rho$ defined by $\rho^6=r^6+a^6$),
\begin{eqnarray}
ds^2=\frac{d\rho^2}{\Big(1-\frac{a^6}{\rho^6}\Big)}+
\frac{\rho^2}{9}\Big(1-\frac{a^6}{\rho^6}\Big)\big(d\psi-\frac{3}{2}\sin^2 
\sigma (d\beta+\cos \theta d\phi)\big)^2 + \hspace{1in}\nonumber \\
\hspace{0.7in}+\rho^2 \Big(d\sigma^2+\frac{1}{4}\sin^2 \sigma 
(d\theta^2+\sin^2 \theta 
d\phi^2)+\frac{1}{4}\sin^2 \sigma \cos^2 \sigma(d\beta+\cos \theta 
d\phi)^2\Big). \label{metricr}
\end{eqnarray}
In what follows, we will move back and forth freely between the radial 
coordinates $r$ and $\rho$.
The form of the metric demonstrates that far away from the 
resolution ($\rho 
\rightarrow \infty$), the angular part of the metric reduces to that of 
the Lens space $S^5/\IZ_3$ thought of as a $U(1)$ fibration over $\IC 
\IP^2$. The second term on the first line in (\ref{metricr}) corresponds 
to the $U(1)$ 
fibration, and the second line corresponds to the $\IC\IP^2$ base. The 
$\IC \IP^2$ metric here is the standard Fubini-Study metric 
(see for instance, equation (5.2) in \cite{Gauntlett:2004yd}). 
It should be emphasized that the ranges of the angles involved 
are important here. To get $\IC\IP^2$ as the base, we need the periodicity 
of $\beta$ to be $4\pi$ and the ranges of $\phi$ and $\theta$ to be 
$2\pi$ and $\pi$ respectively. To make sure the fibration is indeed the 
Lens space, which is what corresponds to the resolution of the orbifold, 
we also need to fix the 
$\psi$-periodicity to be $2\pi$. If instead we took the periodicity of 
$\psi$ to be $6\pi$, we get $S^5$ instead of $S^5/\IZ_3$. To fix the 
range of $\sigma$ we need to look at the defining relations for the 
angles in terms of the $w_i$ given above. This fixes\footnote{Another way 
to fix this 
is to use the fact that when the period of $\psi$ is $6\pi$ we need to 
recover the 5-sphere, whose angle integral we know should be $\pi^3$. 
Using this and the angular part of the measure
(\ref{measure}), we can integrate to find the $\sigma$-range.} 
the range of $\sigma$ to be from $0$ to $\pi/2$. For later use, we also 
write 
down 
$\sqrt{g_6}$ for the resolved $\IC_3/\IZ_3$, 
\eqn{measure}{\sqrt{g_6}=\frac{1}{24} \rho^5  \sin^3 \sigma \cos \sigma 
\sin \theta,}
which happens to be the identical to the unresolved case in the $a 
\rightarrow 0$ limit.

The metric is clearly a higher dimensional generalization of the usual 
Eguchi-Hanson ALE gravitational instanton familiar from four dimensional 
gravity. The 4D Eguchi-Hanson is the analogous metric on the
the resolution of $\IC^2/\IZ_2$.

Before proceeding let us stress again that there is a single parameter $a$, which controls the size
of the ``blown-up" cycle. This parameter corresponds to a dimension six 
operator on the gauge theory side.
This follows from the standard $AdS/CFT$ analysis exactly like in the 
case studied in detail in 
\cite{Benvenuti:2005qb}, where the authors considered a four cycle resolution of the $\IZ_2$-orbifolded 
conifold\footnote{Using the terminology of \cite{Benvenuti:2005qb}
the $a$ parameter corresponds to a ``local" (dimension six) deformation, 
while the ``global" (dimension two)
deformation  does not exist in our case.}.

With this metric, our next task is to calculate the scalar Laplacian so 
that we can do supergravity as sketched in the last section. 
This 
is most easily done in terms of differential forms, using the 
fact that $\Box = *d*d$ where $*$ stands for the Hodge dual. The form of
the metric immediately lets us write it as a sum of squares of  
(non-coordinate) basis forms, and working with them simplifies the 
computation of the Laplacian. This is because the Hodge duals are trivial 
to compute in terms of these, as opposed to the coordinate basis. The 
final result, after the dust settles, is
\begin{eqnarray}
\Box H = \Box_\rho H +\frac{1}{\rho^2}\Box_\sigma H=  
\Big(1-\frac{a^6}{\rho^6}\Big)^{1/2}\partial_{\rho}\left(
\Big(1-\frac{a^6}{\rho^6}\Big)^{1/2}\partial_\rho H\right)+ \nonumber 
\\
+\frac{1}{\rho}\Big(5-\frac{2a^6}{\rho^6}\Big)\partial_\rho H + 
\frac{1}{\rho^2}\big(\partial_\sigma^2 H + 
2(\cot \sigma + \cot 2\sigma)\partial_\sigma H \big). \label{appC}
\end{eqnarray}
The Laplacian here is written under the assumption that 
$H=H(\rho,\sigma)$, so the other angular coordinates drop off. But we do 
present the full Laplacian in an appendix for the viewing pleasure of 
the reader. 

The reason we can get away with looking at only the dependence on one of 
the angles, $\sigma$, is as follows. First, we are placing the stack of 
D3-branes precisely {\em at} the resolution $\rho=a$, where the $U(1)$ 
fibration has shrunk to zero size and therefore the dependence can only 
be on $\rho$ and the coordinates of the blown-up four-cycle. This is 
easily understood through an
analogy in two dimensions. If we keep the source at any point other than the
origin, we expect the Green's function to depend both on the radial 
coordinate
and the polar angle. But if we choose to place the source precisely at 
the origin (where the one-cycle, the circle corresponding to the polar angle, 
has shrunk to zero size), then the Green's function is independent of the  
angle and is purely a function of the radius. The second simplification 
happens because by symmetry, we are free to place the stack on the 
``North pole" of the four-cycle, $\sigma=0$. Once we make that choice, the 
same kind of argument as above applies again, and we have a solution that 
is independent of the rest of the angles of $\IC\IP^2$. In any event, the 
end result is that we can capture all the interesting physics by just 
looking at the $\rho, \sigma$-dependence of the warp factor, $H$. 

If we ignore the $\sigma$-dependence as well, on the other hand, we loose 
some physics, because then we are making the assumption that 
the D-brane sources are smeared over the resolved $\IC\IP^2$ instead of 
being localized. A solution of this form, but for the case of the resolved 
conifold, was constructed by Pando Zayas and Tseytlin \cite{Pando 
Zayas:2000sq}. For the resolved $\IC^3/\IZ_3$ we present a smeared 
solution in Appendix D. This gives us a nice consistency check: the full 
Green's function that we construct in the following should reproduce 
the smeared result, when we look at its singlet-under-$\sigma$ 
part.

The equation that we need to solve for the unsmeared case takes the form:
\eqn{Gre}{\Box H = 
\Box_\rho H+\frac{1}{\rho^2}\Box_\sigma H=-\frac{3C}{4\pi^3\rho^5\sin^3 
\sigma 
\cos 
\sigma}\delta(\rho-a)\delta(\sigma).}
The normalization on the delta function is again fixed by integrating over 
the suppressed angles. For future convenience, we will define
$\sqrt{g_\rho}\equiv \rho^5$ and $\sqrt{g_\sigma}\equiv \sin^3
\sigma \cos \sigma$. The standard technology for solving such 
Poisson-type equations dictates that we
proceed by first solving $ \Box_\sigma Y_l=-E_l 
Y_l$, where the $Y_l$ satisfy   
\begin{eqnarray}
\int Y_l^*(\sigma) Y_{l'}(\sigma)\sqrt{g_\sigma}d\sigma=\delta_{ll'}, 
\label{orth1} \\
\sum_l Y_l^*(\sigma_0) Y_{l}(\sigma) =\frac{1}{\sqrt{g_\sigma}} 
\delta(\sigma-\sigma_0). \label{orth2}
\end{eqnarray}
The next step is to solve the remaining (radial) part, 
\eqn{radial}{\Box_\rho H_l(\rho,\rho_0)-\frac{E_l}{\rho^2}H_l(\rho, \rho_0)= 
-\frac{3C}{4\pi^3\rho^5}\delta(\rho-\rho_0).}
while using
the boundary conditions relevant to the problem.
Then, it is easy to check that 
\eqn{Hgeneric}{H(\rho,\rho_0=a,\sigma,\sigma_0=0)=\sum_l 
H_l(\rho,\rho_0=a)Y_l^*(\sigma_0=0)Y_{l}(\sigma),}
is the desired solution to (\ref{Gre}).

The first step in this program is the solution of the angular part. This 
can in fact be recast as a hypergeometric equation, with the solution
\eqn{angsol}{Y_l=A_l \ {}_2 F_1(-l, l+2, 1, \cos^2 
\sigma) = A_l \
P_{l}^{(0,1)}(-\cos 2\sigma),}
where we re-write the hypergeometric function in terms of an orthogonal 
(Jacobi) polynomial to emphasis the fact that they satisfy an 
orthonormality relation. The orthonormality relation of the Jacobi 
polynomials \cite{Handbook} in the present context takes precisely the form 
we want, i.e. 
eqn(\ref{orth1}), provided we take the normalization to be
\eqn{nrm}{A_l=2\sqrt{l+1}.}
Notice also that 
with the malice of hindsight, we have written 
$E_l=2l(2l+4)$. We mention also that the Jacobi 
polynomials are defined only if we choose $l$ to be an integer. Since the 
angular ``energy" eigen values of the harmonics on the $d$-sphere take 
the form $l(l+d-1)$ (with $d=5$ for our case) for any integer $l$, we are 
missing half of the harmonics in our description. 
Let us now show that 
this is just an immediate result of the fact that we are interested only in the 
$\psi$-independent harmonics, since, as we have already explained,
this angle collapses at the tip leaving only the $\IC\IP^2$. Obviousely, any $S^5$ harmonics 
can be multiplied by the $r^{l(l+4)}$ factor making it a solution of the full $\mathbb{R}_6$
Laplacian equation, which we will denote by $\Psi_l$. 
On the other hand, any such solution can be expressed in terms of the 
$\mathbb{C}_3$ holomorphic coordinates (\ref{eq:w1w2w3}). Morever, this 
dependence on $w_i$'s have to be homogenous, both due to the $r$ 
dependence of $w_i$'s and because by definition $\Psi_l$ is a regular
function of the angles. This means that $\Psi_l$ looks like a sum of 
terms each being a product of exactly $l$ $w_i$'s or $\bar{w_i}$'s. 
Finally, to cancel the $\psi$-dependence of $\Psi_l$ we need equal number 
of $w_i$'s and of $\bar{w_i}$'s in each term, so, indeed, only harmonics 
with even $l$ can contribute to our expansion\footnote{We are grateful to 
the anonymous referee for this elegant explanation.}.

In any event, with the angular harmonics at hand, we turn to the radial 
part which takes the form
\eqn{rad1}{\Big(1-\frac{a^6}{\rho^6}\Big)\frac{d^2 
H_l}{d\rho^2}+\Big(\frac{5}{\rho}+\frac{a^6}{\rho^7}\Big)\frac{d H_l}{d 
\rho}-\frac{2l (2l+4)}{\rho^2}H_l=-\frac{3 
C}{4\pi^3\rho^5}\delta(\rho-a).}
Away from the stack, these can again be solved in terms of hypergeometric 
functions, and the independent solutions are
\begin{eqnarray}
H_l^A&=&{}_2F_1 \Big(\frac{2}{3}+\frac{l}{3}, \ -\frac{l}{3}, \ 
\frac{2}{3}; 
\ \frac{\rho^6}{a^6}\Big), \\
H_l^B&=&\frac{\rho^2}{a^2} \ {}_2F_1 
\Big(1+\frac{l}{3}, \ \frac{1}{3}-\frac{l}{3};\ \frac{4}{3};
\ \frac{\rho^6}{a^6}\Big). 
\end{eqnarray}
The radial equation has  a symmetry under $l \leftrightarrow  
-(l+2)$, which manifests itself in the solutions above as the symmetry of 
the hypergeometric function ${}_2F_1 (a, \ b; \ c; z)$ under $a 
\leftrightarrow b$. This choice can be judiciously used to simplify 
some of the computations below.

We want to look for a specific solution that is defined for all $r>0$ 
($\rho > a$), 
with the condition that it should vanish at infinity. This fixes it upto 
an overall 
constant which in turn we can determine by integrating across the delta 
function. The word 
{\it across} here should be taken with a grain of salt because this is a 
radial delta function 
localized at the (equivalent of the) origin. Effectively this means that 
when we 
integrate (\ref{rad1}), the entire contribution to the discontinuity on the first 
derivative comes from the ``outside" piece because there is no
inside piece at the origin\footnote{To be more precise, we should put the 
delta function away from $r_0=0$ ($\rho_0=a$) and then let $r_0 
\rightarrow 0$ {\it after} the matching, but for sufficiently well-behaved functions, 
this gives the same result as our recipe here.}. The end result is:
\begin{eqnarray}
\label{temp} H_l^{{\rm out}} \equiv C_l H_l^{O} =
C_l\Big[H_l^A-\frac{\Gamma(1+\frac{l}{3})^2\Gamma(-\frac{1}{3})}
{\Gamma(2+\frac{l}{3})^2\Gamma(\frac{1}{3})}H_l^B\Big], \\
{\rm where}  \ \ C_l=\frac{C 
\Gamma(-\frac{l}{3})\Gamma(\frac{l}{3}+\frac{2}{3})}
{4\pi^3a^4\big(3-\sqrt{3}\cot(\frac{\pi l}{3})\big)\Gamma (\frac{2}{3}) 
}.\hspace{0.5in}
\end{eqnarray}
In fixing the normalization above, it is useful to notice that the 
unnormalized warp factor $H_l^O$ behaves near $r=0$ as
\eqn{hlout}{H_l^{O}\longrightarrow 
\frac{3\big(\sqrt{3}\cot(\frac{\pi l}{3})-3\big)\Gamma(\frac{2}{3})}
{\Gamma(\frac{-l}{3})\Gamma(\frac{l}{3}+\frac{2}{3})}\log(r),
}
as can be checked.

Using all the above ingredients, and using (\ref{Hgeneric}), 
we can finally write down the 
warp factor for the stack on the ``North pole" of the resolution as
\eqn{warp}{H(\rho,\sigma)=\sum_{l=0}^{\infty}4(-1)^{l} 
(l+1)^2 
P_{l}^{(0,1)}(-\cos 2\sigma) \
H_l^{{\rm out}}(\rho).}
We have used the fact that $P_l^{(0,1)}(-1)=(-1)^l(l+1)$. 

\subsection{Consistency Checks and Holographic RG Flow}

As already commented, the solution found above should reproduce the 
result of Appendix D, 
when we restrict ourselves to the $l=0$ terms. This is indeed what we 
find:
\begin{eqnarray}
H (\rho,\sigma)|_{(l=0)}=4 H_{l=0}^{{\rm out}}(\rho)
&=&
\frac{16 \pi^2g_s N\alpha'^2}{a^4\sqrt{3}}\Big[1+\frac{(a^6+r^6)^{1/3}}
{a^2\Gamma(\frac{4}{3})\Gamma(\frac{2}{3})} \ {}_2F_1\Big(\frac{1}{3}, 1; 
\frac{4}{3}; 1+\frac{r^6}{a^6} \Big)\Big] \nonumber \\
&=&\frac{12 \pi g_s N\alpha'^2}{r^4} {}_2F_1 \Big(\frac{2}{3}, 
\frac{2}{3};
\frac{5}{3}; -\frac{a^6}{r^6} \Big),
\end{eqnarray}
where, in the last line we have used a functional identity relating 
hypergeometric functions \cite{wolfram1}\footnote{Incidentally, the same 
identity can be used to rewrite (\ref{temp}) as $\sim \frac{1}{r^{4+2l}}  
{}_2 F_1 (\frac{2+l}{3},\frac{2+l}{3};
\frac{5+2l}{3}; -\frac{a^6}{r^6} \Big)$, which has the advantage that 
its convergence at infinity is manifest.}. From the asymptotic of the 
hypergeometric function, it is clear that at $r \rightarrow \infty$ the 
correction from the $\frac{1}{r^4}$ behavior is proportional to $a^6$, 
which ties in with the expectations from \cite{Benvenuti:2005qb}.

It can also be checked that this singularity at $r=0$ arising from the 
smearing is removed by the sum over the various $l$'s. In our case, from 
the small $r$ behavior of $H_l^{{\rm out}}$ presented above, we see that 
this 
sum takes the form 
\begin{eqnarray}
H&=&\frac{-C}{8\pi^3a^4}\log(\rho^6-a^6)\sum_{l=0}^{\infty}4(-1)^l(l+1)^2
P_l^{(0,1)}(-\cos 2\sigma) \nonumber \\
&=&\frac{-C}{8\pi^3a^4}\log(\rho^6-a^6)\frac{\delta(\sigma)}{\sqrt{g_\sigma}},
\end{eqnarray}
where in the last line, we have used the completeness relation 
(\ref{orth2}) 
for Jacobi polynomials \cite{wolfram2}. This makes it immediately clear 
that the 
singularity of in the smeared case at $r=0$ is removed because of the 
vanishing of the delta-function away from the 
location of the stack ($\sigma=0$), in our case. The smearing of the 
source on the four-cycle is also evident because the radial
part takes the form  $\log(\rho^6-a^6) \sim \log r$, which is nothing 
but the Green's function in (the remaining) two dimensions.

We can in fact do more, if we keep track of the $l$-dependence of the 
$H_l^{{\rm out}}$ in the sum. In fact, the sum of all the various 
$l$ pieces near $r=0$ should gives rise to an AdS 
throat, because around a smooth point, all spaces are locally flat. The 
emergence of the throat is most easily 
seen if we approach $r=0$ along 
$\sigma=0$, because then the warp factor looks like
\eqn{wn}{H(r)=\sum_{l=0}^{\infty}4(l+1)^3H_l^{{\rm out}}(r).}
We are interested in the near-horizon behavior where the local curvatures 
have become negligible, which means we are working in the limit where 
the distance scales are much less than the resolution size, $r \ll a$. We 
can solve the radial equation (away from the source) in this limit, and
the solution that dies down at infinity can be expressed in terms of 
modified Bessel functions of the second kind \cite{wiki1}. The details of 
the solution 
are irrelevant to us, except for one piece of information: the entire 
dependence of the 
solution on $r$ and $l$ (the normalization fixed by integrating 
across the delta function turns out to be independent of $l$), is captured 
by the combination 
$\sqrt{l(l+2)}\ r^3 \sim l \ r^3$.  So we can write
\eqn{www1}{H(r) \sim \sum_l^{\infty} l^3 f(lr^3),}
where the fact that we expect this sum to be convergent in $l$ implies 
that the function $f(lr^3)$ can be thought of as a regulator\footnote{This 
clever trick is taken straight from Klebanov-Murugan 
\cite{Klebanov:2007us}.}. In particular, the regulator accomplishes 
finiteness by decaying rapidly for $l > \frac{1}{r^3}$, so we can restrict 
the sum as
\eqn{www2}{H(r) \sim \sum_{l=0}^{1/r^3}l^3 f(lr^3) \sim \int_0^{1/r^3} 
l^3 f(lr^3)dl\sim\frac{1}{r^{12}}\int_0^1 x^3f(x)dx = \frac{{\rm 
const.}}{r^{12}}.}
The second to last step involves a change of variables, and it can be 
checked that the final integral converges for the modified Bessel 
function mentioned earlier: for small $r$, it can be approximated by  
$\log r$. All that remains, is to notice that close to $\rho=a$ (with 
$\sigma =0$), the metric (\ref{metricr}) takes the flat form with a 
new radial coordinate $u=\frac{r^3}{3a^3}$. So in terms of this flat 
coordinate, the 
warp factor goes as $\sim \frac{1}{u^4}$, suggesting the emergence of the 
AdS throat through the usual arguments.

\section{\bf RG Flow in the Gauge Theory}
\label{five}

In this section we will describe various RG flows triggered by non-zero VEVs of the fields $A_i$, $B_i$ and $C_i$.

We will start with the unresolved case. The supergravity counterpart of 
this discussion can be found in Appendix C. The unresolved case 
corresponds to $t_a=0$ in 
(\ref{FI}).
As we have already explained above the set (\ref{FI}) does not describe the moduli space, since we also have
to impose the F-terms conditions. For instance, it immediately follows 
that setting $A_1=B_2=C_3=v$
and giving zero VEVs to the rest of the fields
is consistent with the D-terms equations, but still does no satisfy the F-term restrictions,
since $A_1B_2C_3 \neq A_2B_3C_1=0$.
The VEVs that do not contradict the F-term relations are of the form $A_3=B_3=C_3=v$.
We see that in this case $U_{333}=v^3$
and all the other mesons are zero, so the VEVs corresponds to a point on the unresolved orbifold away from the apex.
To be more precise, comparing (\ref{homog}) and (\ref{MO}) we learn that
$\left \vert z_3 \right \vert= \frac{1}{\sqrt{3}}\left\vert z_4 \right\vert=v^{3/4}$.

Let us now analyse the superpotential.
The $SU(N)_1 \times SU(N)_2 \times SU(N)_3$ gauge group is broken down to a single $SU(N)$.
Substituting  $A_3=B_3=C_3=v$ into (\ref{sup}) we find that 
the mass matrix for the remaining six fields 
has rank four\footnote{We would like to thank the referee for pointing out to us an obvious mistake 
regarding the matrix rank
in the earlier version
of the paper.}.
Being more specific, the eigenvalues are $0$, $\frac{\sqrt{3}}{2} v$ and $-\frac{\sqrt{3}}{2} v$,
each having degeneracy two.
The matrix appears in Appendix E, where we also report its related eigenvectors.
Since out of three $SU(N)$ gauge groups two are broken by the VEVs, two chiral superfields
should be ``eaten" by the corresponding vector multiplets to form two massive multiplets.
This is easily seen in the unitary gauge. Indeed, the gauge is
completely fixed by putting $A_3=B_3=v$, while we still have to take into account the fluctuations of $C_3$. 
As for the rest of the fields the following parameterization proves to be
convenient:
\bea
&& C_3 = C_3 + \Phi, \quad A_1 = \Phi_1 + 2 \Psi_1, \quad  B_1 = \Phi_1 - \Psi_1 - \Psi_4, \quad C_1 = \Phi_1 - \Psi_1 + \Psi_4, \non \\
&& \qquad \qquad A_2 = \Phi_2 + 2 \Psi_3, \quad  B_2 = \Phi_2 + \Psi_2 - \Psi_3, \quad C_2 = \Phi_2 - \Psi_2 - \Psi_3.
\eea
Substituting this into the superpotential we find:
\bea
\label{eq:W3}
W &=& 6 v \Tr (\Psi_1 \Psi_2 + \Psi_3 \Psi_4) + \Tr \left( \Phi \left[ \Phi_1, \Phi_2 \right] \right) - \non \\
&& - \Tr \bigg( \Phi \Big( \{ \Phi_1, \Psi_2 \} + \{ \Phi_2, \Psi_4 \} 
                             + [\Phi_1, \Psi_3] - [\Phi_2, \Psi_1]  + \non \\
&&  \qquad + \{\Psi_1, \Psi_2\}  +  \{\Psi_3, \Psi_4\}  + [\Psi_1,\Psi_3] - [\Psi_2,\Psi_4] \Big)  \bigg).
\eea
As expected four fields, $\Psi_1$, $\Psi_2$,$\Psi_3$, and $\Psi_4$, 
become massive and we have to integrate all of them out.  
As a result all the terms in the last two lines of the superpotential expression become quartic in terms of the
three surviving fields $\Phi$, $\Phi_1$ and $\Phi_2$. The quartic terms will be irrelevant in the IR, thus we are left only with the second term of (\ref{eq:W3}), which is precisely the $\N = 4$ superpotential\footnote{Alternatively one can notice that the fields $\Psi_i$ have dimension two, so all the operators in the last two lines of (\ref{eq:W3}) are IR irrelevant.}.

We now consider the resolved orbifold case. For D3-branes localized away from the resolved apex, the
RG flow is essentially the same as for the unresolved case. The only difference is that the VEVs
of $A_3$, $B_3$ and $C_3$ cannot anymore be equal. Let us focus instead on D3-branes located at the point
$\left\vert z_3\right\vert^2 = \mu$ of $\IC\IP^2$.  

In this case, we found that there are two different scenarios that describe the RG flow.
According to the first, two bi-fundamentals acquire VEVs (say 
$A_3=B_3=v$), while in the 
second only one VEV is non-trivial (without losing generality
we will put $C_3=v$). 
Notice that in both cases all the mesonic fields vanish and the $SU(3)$ symmetry is broken down to $SU(2)$
matching the geometry expectations. 
Unfortunately, since we don't know the precise form of the dual dimension six operator, we cannot directly check
what scenario is the right one. Instead, we will describe both possibilities arguing that in any case
the RG flow re-produces the $\N = 4$ field content and the superpotential.

\subsection{The $A_3=B_3=v$ RG flow}

Again, these VEVs leave only one un-broken $SU(N)$. Furthermore, plugging 
the VEVs into the superpotential
we find that the fields $C_1$ and $C_2$ become massive, so we have to integrate them out.
This, in turn, leads to two constraints: $A_1=B_1$ and $A_2=B_2$. 
Substituting this into the remaining two terms in the superpotential 
we immediately arrive at the $\N = 4$ superpotential:
\eqn{W0}{W = {\rm Tr} \left( C_3 \left[A_1, A_2 \right] \right). }

\subsection{The $C_3=v$ RG flow} 

\begin{figure}
\begin{center}
\includegraphics[width=0.9\textwidth]{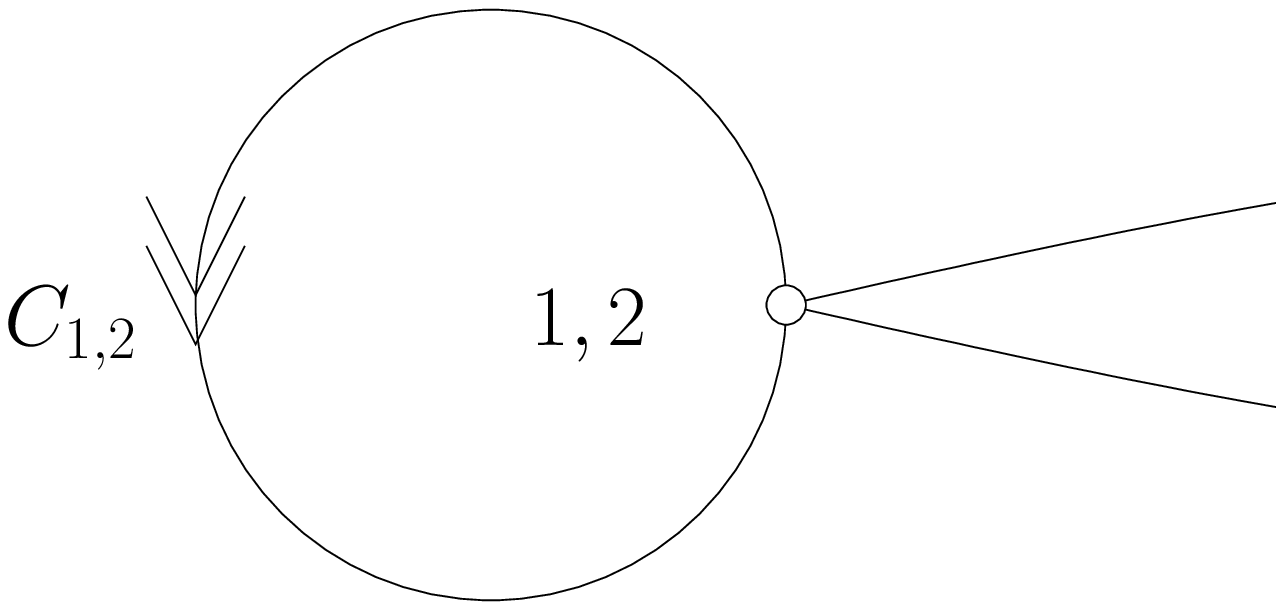}
\caption{The gauge theory quiver for the theory after the first ``step" of 
the RG flow. The VEV of the field $C_3$
breaks $SU(N)_1 \times SU(N)_2$ down to $SU(N)_{1,2}$ and the massive fields $A_{1,2}$ and $B_{1,2}$
are integrated out. At the second step, node 3 confines.}
\label{Quiver2}
\end{center}
\end{figure}

Now the superpotential reads\footnote{For the sake of 
simplicity we omit here the group traces and indices.}:
\eqn{W1}{W = v A_1 B_2- v A_2 B_1 + A_2 B_3 C_1 - A_1 B_3 C_2 + A_3 B_1 C_2 - A_3 B_2 C_1 }
Integrating the massive fields $A_{1,2}$ and $B_{1,2}$ out we arrive at the following result:
\eqn{W2}{W = \frac{1}{v} {\rm Tr} \left(A_3 B_3 \left[ C_1, C_2 \right] \right). }
The VEV of $C_3$ breaks also $SU(N)_1 \times SU(N)_2$ down to the diagonal 
$SU(N)$
so we end up with the quiver depicted in Fig. (\ref{Quiver2}).
This theory, however, is not conformal anymore, since we now have $N_{\rm f}=N_{\rm c}$ for the 3d node on the quiver
on Fig. (\ref{Quiver2}). The theory flows to the strongly coupled region and the right node confines.
Apart from the adjoint fields $C_1$ and $C_2$ we have a meson $M=A_3 B_3$, 
which also
transforms in the adjoint of the left node of the quiver diagram. The field content, therefore, exactly reproduces
the ${\cal N}=4$ SYM theory. The superpotential, however, involves an additional non-perturbative Affleck-Dine-Seiberg term:
\eqn{W3}{W = \frac{1}{v} {\rm Tr} \left(M \left[ C_1, C_2 \right] \right) + X( {\rm det} M - B \tilde{B} - \Lambda^{2N}).}
Here $B$ and $\tilde{B}$ are the baryon fields, $\Lambda$ is the strong coupling scale and the chiral field $X$
is the Lagrange multiplier imposing the quantum moduli constraint in the parentheses.
In the deformed conifold model this constraint exhibits two completely
\emph{separated} branches of solutions \cite{Klebanov:2000nc, Klebanov:2000hb},  
since the meson and the 
baryons cannot acquire VEVs
simultaneousely. In the present model the two branches are not separated. This, however,  does not modify the observation that the superpotential 
is still cubic. For instance, the ``purely" mesonic branch corresponds to
$B=\tilde{B}=0$ and ${\rm det} M = \Lambda^{2N}$, which can solved by
$M=\mathbf{1}_{N \times N} \Lambda^2$. Substituting this into (\ref{W3}) we find that the superpotential vanishes,
so, similar to what we did in the unresolved case, we have to consider next to the leading order corrections.
We will set:
\eqn{}{M=\mathbf{1}_{N \times N} \Lambda^2+\frac{\Phi}{v},}
reproducing eventually the ${\cal N}=4$ superpotential:
\eqn{}{W = {\rm Tr} \left( \Phi \left[C_1, C_2 \right] \right).}

\section{\bf Discussion}

In this paper we have constructed a supergravity solution that describes 
an RG flow from  the gauge theory dual to $AdS_5 \times S^5/\IZ_3$, to  
${\cal N}=4$ SYM theory which in turn is dual to $AdS_5 \times S^5$.  The
resolution of the $\IZ_3$ singularity is given by a $\IC\IP^2$ placed at 
the tip of the cone.
From the explicit metric that we wrote down, it can be checked 
\cite{Benvenuti:2005qb} 
that far away in the UV, the leading order correction from AdS describes 
an order six operator acquiring VEV in the gauge theory. This VEV is 
nothing but the resolution parameter in the geometry. The explicit form of 
this operator unfortunately is unknown.

The problem of constructing a 10$d$ supergravity background based on the 
resolved 6$d$ space reduces to
finding the six-dimensional Green function, which serves as a warp 
function in the full solution.
If the function, however, is taken to depend only on the
radial coordinate, and so the D$3$-brane source is ``smeared" over the 
$\IC\IP^2$,
the total 10$d$ solution proves to be
singular. This is just a common phenomenon for 10$d$ backrounds based on 
``smeared" D-branes\footnote{See \cite{Pando Zayas:2000sq} for the 
resolved conifold example.}.
To avoid the singularity we have placed the D$3$ branes stuck at a point 
of the $\IC\IP^2$. 
The map between the D$3$ branes coordinates and the gauge invariant 
(mesonic) combination of the fields was presented in Section \ref{gastr}.

Back in the dual
gauge theory, this corresponds to giving VEVs  to some  bi-fundamental 
fields. 
Although the precise form of the dimension six operator dual to the 
resolution parameter
remains a mystery, we are able to show that 
there are only two distinct ways in which the fields can acquire non-trivial VEVs.
In both cases the non-trivial VEVs break the $SU(3)$ isometry, the conformal
invariance of the original theory and trigger (presumably) different RG 
flows to the same $\N=4$ SYM theory.

We have argued that in the second scenario the RG flow proceeds in two ``steps". First, some 
fields acquire VEVs breaking
three gauge groups down to two. Integrating the massive fields out, we 
are left with four
(two adjoints and two bi-fundamentals) out of nine original fields and 
a \emph{quartic} superpotential. Second,
one of the nodes confines. In terms of the two adjoint fields
and a new meson constructed from the two bi-fundamentals, the 
superpotential becomes \emph{cubic} like in the ${\cal N}=4$
SYM theory.
It is very tempting, therefore, to identify these two steps on the 
supergravity side.
Unfortunately, it does not look possible if one wants to stick with the 
supergravity approximation,
which requires large t'Hooft coupling $\lambda = N g_{YM}^2$.
Indeed, the strong coupling scale on the second ``step" of the RG flow is 
related to the energy scale $E$
set by the VEV $v$ as:
\eqn{}{\Lambda = E e^{-\frac{8 \pi^2}{2 N g_{YM}^2(E)}}  \qquad {\rm 
where} \qquad E=\sqrt{v}.}
We see that, for large $\lambda = N g_{YM}^2$ the scale $\Lambda$ and $E$ 
are 
of the same order
and we cannot distinguish between them\footnote{We are grateful to 
Riccardo Argurio for explaining this subtlety.}.
  This situation is actually familiar from the conifold model cascade 
\cite{Klebanov:2000hb}, where
on the supergravity side the cascade steps are ``smoothed" out for the 
same 
reasons as in our case.

A number of interesting questions can be raised for further investigation.
In \cite{Klebanov:2007us}  the presence of the baryonic VEV was verified 
from the analysis of the DBI action
of an Euclidian D$3$-brane wrapped on a 4$d$ cycle of the resolved 
conifold.
Similar computation can also be performed in our case. 
Remarkably, from the gauge theory point of view 
this condensate has to be the same in both suggested RG flow scenarios. Indeed, the baryon constructed from the 
field $A_3 B_3$ and the baryon built from $C_3$ describe the same 
baryonic operator, since their
product can be written only in terms of the meson $A_3 B_3C_3$. This, and 
the fact that the Fayet-Iliopoulos parameters (\ref{FI}-\ref{FI3}) turn 
out to be 
identical for both these choices, leads us to speculate that perhaps the 
two 
RG flows 
are dual descriptions of each other.

Finally, it will be very interesting to find the dimension six operator dual to the resolution parameter $a$.
We hope that our paper might provide useful results towards this 
direction, along the lines of \cite{Benvenuti:2005qb}.

\section{Acknowledgments}

We would like to thank Riccardo Argurio, Cyril Closset and Carlo 
Maccaferri for helpful discussions. 
This work is supported in part by IISN - Belgium (convention 
4.4505.86), by the Belgian National
Lottery, by the
European Commission FP6 RTN programme MRTN-CT-2004-005104 in which the 
authors
are associated with V. U. Brussel, and by the Belgian Federal Science
Policy Office through the Interuniversity Attraction Pole P5/27.

\section*{\bf Appendix}
\addcontentsline{toc}{section}{Appendix}

\subsection*{{\bf A.} \ Metric on the ALE space}
\addcontentsline{toc}{subsection}{{\bf A} \ Metric on the ALE space}
\renewcommand{\theequation}{A.\arabic{equation}}

One way to derive the metric on the resolution of $\IC^3/\IZ_3$ is to use 
the fact that the K\"ahler form
$K(r)$ on this space depends only on $r^2=|w_1|^2+|w_2|^2+|w_3|^2$.
Since the space is Calabi-Yau, among other things, it is both
Rici-flat and K\"ahler. So the metric can be written as $g_{i\bar
j}=\partial_i \partial_{\bar j} K$,
and then the Ricci-flatness condition turns out to be a 
differential equation for $K(r)$:
\begin{eqnarray}
{\rm det}(\partial_i\partial_{\bar j}K)={\rm const.}
\end{eqnarray}
After absorbing the irrelevant constant by rescaling $K(r)$, this
translates to
\eqn{K}{(K')^2(r^2K''+K')=1.}
The primes here are with respect to $r^2$. It will prove convenient to
introduce a new function ${\cal F}$ defined by
\eqn{F}{{\cal F}\equiv r^2 K',}
in terms of which the differential equation above has the simple solution
\eqn{F1}{{\cal F}=(r^6+a^6)^{1/3},}
with $a^6$ an integration constant which reflects the resolution of the 
space. By tuning $a$ to zero, we can recover the unresolved space. It is 
easy to integrate ${\cal F}$ once again to obtain an explicit expression 
for $K(r)$ in terms of hypergeometric functions, but we will not need it, 
so we will refrain from writing it down. 

Using these, and the angular variables defined in section 4 in terms of 
the $w_i$, we can calculate the metric directly as $ds^2=g_{i\bar j}dw^i 
d{\bar w}^{\bar j}$. The result is 
\begin{eqnarray}
ds^2={\cal F}'dr^2+\frac{{\cal 
F}'r^2}{9}\big(d\psi-\frac{3}{2}\sin^2
\sigma (d\beta+\cos \theta d\phi)\big)^2 + \hspace{1.7in}\nonumber \\
\hspace{0.6in}+{\cal F} \Big(d\sigma^2+\frac{1}{4}\sin^2 \sigma
(d\theta^2+\sin^2 \theta
d\phi^2)+\frac{1}{4}\sin^2 \sigma \cos^2 \sigma(d\beta+\cos \theta
d\phi)^2\Big).
\end{eqnarray} 
This turns into the metric presented in Section \ref{resolvedbrane}, once 
we make 
the 
definition $\rho^2 \equiv {\cal F}$.	

\subsection*{{\bf B.} \ The Scalar Laplacian}
\addcontentsline{toc}{subsection}{{\bf B} \ The Scalar Laplacian}
\renewcommand{\theequation}{B.\arabic{equation}}

In this appendix we exhibit the full Laplacian for the resolved 
$\IC^3/\IZ_3$ space. It takes the form 
\begin{eqnarray}
\Box H = \Big(1-\frac{a^6}{\rho^6}\Big)^{1/2}
\partial_{\rho}\Big(\big(1-\frac{a^6}{\rho^6}\big)^{1/2}
\partial_\rho H\Big)+
\frac{1}{\rho}\Big(5-\frac{2a^6}{\rho^6}\Big)\partial_\rho H + 
\hspace{0.5in}\nonumber \\
+\frac{1}{\rho^2}\Big[\big(\partial_\sigma^2 H + 
2(\cot \sigma + \cot 2\sigma)\partial_\sigma H \big)+\frac{4}{\sin^2 
\sigma \sin \theta}\partial_\theta(\sin \theta \partial_\theta H)+ 
\hspace{0.2in}
\nonumber \\
+\frac{4}{\sin^2\sigma}\Big(\frac{1}{\sin \theta}\partial_\phi -\cot 
\theta\partial_\beta \Big)^2H+\frac{4}{\cos^2\sigma}\Big(\frac{3}{2}\sin 
\sigma
\partial_\psi +\frac{1}{\sin \sigma}\partial_\beta \Big)^2H + \nonumber\\
+ 9\Big(1-\frac{a^6}{\rho^6}\Big)^{-1}\partial_\psi^2 H\Big]. \hspace{2in}
\end{eqnarray}

\subsection*{{\bf C.} \ Branes on the Unresolved $\IC^3/\IZ_3$: Away from
the Tip}
\addcontentsline{toc}{subsection}{{\bf C} \ Branes on the Unresolved
$\IC^3/\IZ_3$: Away from the Tip}
\renewcommand{\theequation}{C.\arabic{equation}}

Here we compute the Green's function for a D-brane stack on 
the unresolved orbifold, but away from the tip ($\rho \ne a$). Far away 
from the stack, we expect 
to reproduce the behavior we calculated in section 3.1, but we also 
expect to see the AdS throat in the near-horizon region because the stack 
is no longer at a singular point. The fact that the space is unresolved 
will be reflected in the fact that the solution is still singular.

To simplify the computations, we will look at the case where the stack is 
at the point $\rho=\rho_0, \ \sigma=0$. The location 
$\sigma=0$ kills the dependence of the Green function on the other angles 
of the $\IC\IP^2$ as was argued in the main body of the paper. 
We will only be interested in the case where the solution is of the form  
$H(\rho,\sigma)$, which means that we are assuming that the D3-branes are 
smeared over the $\psi$-circle.\footnote{It turns out that the more 
general case of $\psi$-dependent warp factor can also be solved exactly in 
terms of Jacobi polynomials (this time, with more general quantum numbers 
than what we saw in Section 4), but we will not present the solution 
here.}
The form of the 
Laplacian permits such a choice. The equation to be solved takes the form
\begin{eqnarray}
\Box_\rho H+\frac{1}{\rho^2}\Box_{\sigma}H=\frac{-3C}{4\pi^3\rho^5\sin^3 
\sigma \cos \sigma}\delta(\rho-\rho_0)\delta(\sigma),\\ 
{\rm where}\ \  \Box_\rho=\frac{1}{\rho^5}\frac{\partial}{\partial 
\rho}\Big(\rho^5\frac{\partial}{\partial \rho}\Big),\hspace{0.75in}
\ \ \nonumber   
\end{eqnarray}
with $\Box_\sigma$ as defined by (\ref{appC}). As before, we now 
solve 
\begin{eqnarray}
&\Big(\Box_\sigma+2l (2l+4) 
\Big)\Sigma(\sigma)=0,& 
\end{eqnarray}
for the angular part in terms of Jacobi polynomials.
The full normalized angular solutions are
\eqn{ylm}{Y_{l}(\sigma)=\sqrt{4(l+1)}
P_{l}^{(0,1)}(-\cos 2\sigma)
.}

The remaining radial part of the equation looks like 
\eqn{apprad}{\frac{\partial^2 H_l}{\partial 
\rho^2}+\frac{5}{\rho}\frac{\partial H_l}{\partial \rho}-\frac{2l 
(2l+4)}{\rho^2} H_l=\frac{-3C}{4\pi^3\rho^5}\delta(\rho-\rho_0),}
whose solution, after matching the function and its derivative through 
the delta function is
\begin{eqnarray}
H_l(\rho,\rho_0)=
\left\{ \begin{array}{ll}
  \displaystyle{\frac{3C}{16\pi^3(l+1)} \rho_0^{-(2l+4)}\rho^{2l} }& 
\ \ \rho 
\leq 
\rho_0, \\
 \\
  \displaystyle{\frac{3C}{16\pi^3(l+1)} \rho^{-(2l+4)}\rho_0^{2l} } 
& \ \ \rho 
\geq 
\rho_0.\\
\end{array} \right.
\end{eqnarray}

The full solution looks like
\eqn{fin}{H(\rho,\rho_0;\sigma,0)=\sum_{l} 
H_l(\rho,\rho_0)Y_{l}^*(0)Y_{l}(\sigma).}
Restricting to $l=0$ gives the dependence far away from the source
\eqn{faraway}{H(\rho) \rightarrow \frac{3C}{16\pi^3 \rho^4}\times 
\big(\sqrt{4} \ \big)^2=\frac{12\pi g_s N\alpha'^2}{\rho^4}}
which reproduces the result obtained in section 3.1, eqn. (\ref{warp1}), 
where the stack was 
assumed to be at the tip. The emergence of the AdS throat close to the 
stack is entirely 
analogous to the resolved case, which was discussed in the main text.

\subsection*{{\bf D.} \ Warp Factor with Smeared Sources}
\addcontentsline{toc}{subsection}{{\bf D} \ Warp Factor with Smeared 
Sources}
\renewcommand{\theequation}{D.\arabic{equation}}

In this appendix, we compute the warp factor for the 
resolved $\IC^3/\IZ_3$, but with the assumption that the source branes are 
smeared all 
over the resolution and there is no angular dependence. We start with 
the Laplacian as defined by
\eqn{boxc}{\Box H = \frac{1}{ \sqrt{g_6} } \ \partial_{\mu} ( \sqrt{g_6} 
g^{\mu\nu} \partial_{\nu} H) = \frac{1}{\rho^5}\partial_\rho
\Big(\rho^5\big(1-\frac{a^6}{\rho^6}\big)\partial_\rho H\Big) 
} 
where we have used (\ref{measure}) and assumed $H\equiv 
H(\rho)$, which is the basis of the smeared source approximation. For 
comparison with the unsmeared solution in the main body 
of the paper, we will solve the Green's equation in terms of  
$r=(\rho^6-a^6)^{1/6}$. The equation is effectively first 
order and can be solved by direct integration in terms of elementary 
functions. But we will write its solution in a 
hypergeometric form 
\eqn{2}{H =
\frac{12\pi g_s N \alpha'^2}{r^4}\ {}_2F_1\Big(\frac{2}{3},\ 
\frac{2}{3}; \ \frac{5}{3};-\frac{a^6}{r^6} \Big),}
for ease of comparison. We have fixed the overall normalization by 
demanding that the warp-function should look like that of the unresolved 
case (\ref{warp1}), far away from 
the resolution ($\frac{a}{r} \rightarrow 0$).

\subsection*{{\bf E.} \ The Mass Matrix for the Unresolved Case}
\addcontentsline{toc}{subsection}{{\bf E} \ The Mass Matrix for the 
Unresolved Case}
\renewcommand{\theequation}{E.\arabic{equation}}

We have argued above that the VEVs $A_3=B_3=C_3=v$ render the 
superpotential quadratic in terms of 
the remaining fields and the symmetric mass matrix $\mathbf{M}$ has rank four.
In this appendix we present the basis in which the mass matrix acquires a diagonal form.
In the $(A_1,A_2,B_1,B_2,C_1,C_2)$ basis we have:

\begin{equation}
\mathbf{M} = \frac{v}{2} \left(
\begin{array}{cccccc}
0 &  0 &  0 &  0 &  1 & -1  \nonumber \\
0 &  0 &  0 & -1 &  0 &  1  \nonumber \\
0 &  0 &  0 &  1 & -1 &  0  \nonumber \\
0 & -1 &  1 &  0 &  0 &  0  \nonumber \\
1 &  0 & -1 &  0 &  0 &  0  \nonumber \\
-1&  1 &  0 &  0 &  0 &  0 
\end{array}
\right) .
\end{equation}
The matrix eigenvalues are $0$, $\frac{\sqrt{3}}{2} v$ and $ -\frac{\sqrt{3}}{2} v$.  Each of the eigenvalues
has degeneracy two.
The corresponding eigenvectors are as follows: 

\beq
\left(   \begin{array}{c}   \frac{1}{\sqrt{3}} \\ \frac{1}{\sqrt{3}}  \\ \frac{1}{\sqrt{3}}  \\ 0 \\ 0 \\ 0 \end{array} \right)
\quad
\left(   \begin{array}{c} 0 \\ 0 \\ 0 \\ \frac{1}{\sqrt{3}} \\ \frac{1}{\sqrt{3}} \\ \frac{1}{\sqrt{3}} \end{array} \right)
\quad
\left(   \begin{array}{c} \frac{1}{\sqrt{3}}  \\ -\frac{1}{2 \sqrt{3}} \\ -\frac{1}{2 \sqrt{3}} 
                                                     \\ 0 \\ \frac{1}{2} \\ -\frac{1}{2}  \end{array} \right)
\quad
\left(   \begin{array}{c} 0 \\ -\frac{1}{2} \\ \frac{1}{2} \\ \frac{1}{\sqrt{3}} \\ -\frac{1}{2 \sqrt{3}} 
                                                               \\  -\frac{1}{2 \sqrt{3}}  \end{array} \right)
\quad
\left(   \begin{array}{c} \frac{1}{\sqrt{3}}  \\ -\frac{1}{2 \sqrt{3}} \\ -\frac{1}{2 \sqrt{3}} 
                                                     \\ 0 \\ -\frac{1}{2} \\ \frac{1}{2}  \end{array} \right)
\quad
\left(   \begin{array}{c} 0 \\ \frac{1}{2} \\ -\frac{1}{2} \\ \frac{1}{\sqrt{3}} \\ -\frac{1}{2 \sqrt{3}} 
                                                               \\  -\frac{1}{2 \sqrt{3}}  \end{array} \right)
\eeq

\newpage

\end{document}